\documentclass[11pt]{article}
\usepackage{amssymb}
\usepackage{amsmath}
\usepackage{amscd}
\usepackage{latexsym,verbatim}
\usepackage[all]{xy}
\xyoption{curve}
\xyoption{line}
\usepackage{graphicx}
\usepackage{graphics}
\usepackage{color}
\def\ii{{\,{\rm i}\,}}

\catcode`@=11
\renewcommand{\section}
{\@startsection{section}{1}{0pt}{\medskipamount}{\medskipamount}{\large\bf}}
\makeatletter\renewcommand{\subsection}
{\@startsection{subsection}{2}{\z@}{-3.25ex plus -1ex minus -.2ex}
{1.5ex plus .2ex}{\it }}

\numberwithin{equation}{section}
\catcode`@=12

\def\e{\epsilon}

\def\cp2star{\star}

\def\beq{\begin{equation}}
\def\eeq{\end{equation}}
\def\bea{\begin{eqnarray}}
\def\eea{\end{eqnarray}}

\renewcommand{\e}{\,\mathrm{e}\,}

\newcommand{\R}{{\mathbb{R}}}

\newcommand{\C}{{\mathbb{C}}}

\newcommand{\Tr}{{\rm Tr}}

\newcommand{\mbf}[1]{{\boldsymbol {#1} }}

\def\>{\rangle}
\def\<{\langle}
\def\+{\dagger}
\def\={\ =\ }

\begin{document}

\begin{titlepage}
\setcounter{page}{0}
\begin{flushright}
HWM--09--5\\
EMPG--09--10\\
\end{flushright}

\vskip 1.8cm

\begin{center}

{\Large\bf Quantum Gravity, Field Theory \\[10pt] and Signatures of
  Noncommutative Spacetime}\footnote{Based on Plenary Lecture
  delivered at the {\sl XXIX Encontro Nacional de F\'{\i}sica de
    Part\'{\i}culas e Campos}, S\~ao Louren\c{c}o, Brasil, September
  22--26, 2008.}

\vspace{15mm}

{\large Richard J. Szabo}
\\[5mm]
\noindent {\em Department of Mathematics\\ Heriot-Watt
  University\\ Colin Maclaurin Building, Riccarton, Edinburgh EH14
  4AS, U.K. \\ and Maxwell Institute for Mathematical Sciences,
  Edinburgh, U.K.}
\\[5mm]
{Email: {\tt R.J.Szabo@ma.hw.ac.uk}}

\vspace{25mm}

\begin{abstract}
\noindent
A pedagogical introduction to some of the main ideas and results of 
field theories on quantized spacetimes is presented, with emphasis on
what such field theories may teach us about the problem of quantizing
gravity. We examine to what extent noncommutative gauge theories may
be regarded as gauge theories of gravity. UV/IR mixing is explained in
detail and we describe its relations to renormalization, to
gravitational dynamics, and to deformed dispersion relations in models
of quantum spacetime of interest in string theory and in doubly
special relativity. We also discuss some potential experimental probes
of spacetime noncommutativity.
\end{abstract}
\end{center}
\end{titlepage}
 
\bigskip

\tableofcontents

\vspace{20mm}

\section{Introduction\label{Intro}}

\noindent
The validity of certain models or hypotheses in theoretical physics is
sometimes argued on the basis of length scales. In modern theoretical
physics, physical phenomena occur from down at the fundamental Planck
scale
\beq
{ \ell_{\rm P}\=\sqrt{{\hbar\,G}/{c^3}}~\simeq~1.6\times
10^{-33}~\mbox{cm} }
\eeq
all the way up to the radius of the observable universe 
\beq
\ell_{\rm universe}~\simeq~ 
4.4\times10^{24}~\mbox{cm}~\simeq~
\big(2.7\times10^{61}\big)\,\ell_{\rm P} \ . 
\eeq
We know that quantum field theory works well in describing the
pertinent physics at least down to the LHC scale
\beq
\ell_{\rm LHC}~\simeq~2\times10^{-18}~\mbox{cm} \ .
\eeq
What happens below this scale, but perhaps still in the regime above
the Planck scale, is a question that will not be answered
satisfactorily by experiment for many years to come. In the passing
time, theoretical physicists may reap the pastures of this unknown
regime and speculate on the mathematical foundations that the correct
modification of quantum field theory might assume at such length scales.

Such is the speculative nature of the notion of ``spacetime
quantization''. In this somewhat radical scenario, the local
coordinates $x^\mu$ are promoted to hermitean operators satisfying
spacetime noncommutativity commutator relations
\beq
{ \big[x^\mu\,,\,x^\nu\big]\=\ii\theta^{\mu\nu} }
\label{spacetimencrels}\eeq
where $\theta^{\mu\nu}$ is a real antisymmetric matrix of dimension
(length)$^2$. If this matrix is constant, then the commutators
(\ref{spacetimencrels}) generate a Heisenberg algebra and imply
spacetime uncertainty relations
\beq
\Delta x^\mu\,\Delta
  x^\nu~\geq~\mbox{$\frac12$}\, \big|\theta^{\mu\nu}\big| \ .
\eeq
This is derived in the same standard way that one would derive
Heisenberg's uncertainty principle from canonical commutation
relations between coordinates $x^\mu$ and momenta $p_\nu$ in quantum
mechanics. In this sense the dimensionful noncommutativity parameters
${ \theta^{\mu\nu} }$ are analogous to Planck's constant {$\hbar$} in
the phase space quantization relations
\beq
\big[x^\mu\,,\,p_\nu\big]\=\ii\hbar\,\delta^\mu{}_\nu \ .
\label{phasespquant}\eeq
In a quantum phase space, points no longer exist and are replaced with
Planck cells of size $\hbar$. von~Neumann thus dubbed the study of
geometrical properties of quantum mechanics as ``pointless''
geometry. In modern parlance of quantized spacetime, this branch of
mathematics has come to be known as ``noncommutative geometry''. The
points of a quantized spacetime become ``fuzzy'' and are replaced with
cells whose size is set by the noncommutativity length scale
$\ell\simeq\sqrt{\theta}$. In general, the tensor $\theta^{\mu\nu}$
can depend on the spacetime coordinates, and even on the momenta (in
which case the algebra of canonical commutation relations is replaced
with an algebra of pseudo-differential operators). We shall see some
explicit examples in what follows.

The proposal of spacetime quantization arises in two major problems of
modern high-energy physics. The first is the problem of
\emph{renormalization}. Just as the Heisenberg uncertainty principle
enables one to avoid the ultraviolet catastrophe in quantum mechanics,
the replacement of points with spacetime cells is a means in which to
tame the ultraviolet divergences of quantum field theory, as an
elegant symmetry preserving alternative to lattice or cutoff
regularizations. This turns out to be an extremely subtle issue and
will be discussed in detail in the following. The second is the
problem of \emph{quantum gravity}. It has been long suspected that
classical general relativity breaks down at the Planck scale ${
  \ell_{\rm P} }$, where quantum gravitational effects become
important. In particular, the classical Riemannian geometry of
spacetime must be replaced by some other mathematical
framework. Since, according to Einstein's theory, gravity affects
spacetime geometry, \emph{quantum} gravity should \emph{quantize}
spacetime. The precise manner in which this quantization occurs is of
course not entirely understood and is one of the biggest challenges
facing modern theoretical physics. Noncommutative geometry may in this
way provide at least some guidance as to how to handle spacetime
structure at very short distances.

In what follows we shall argue that these two fundamental problems are
in fact {\it related} to each other within a systematic and unified
framework of field theory on quantized spacetimes. This theoretical
framework is called {\it noncommutative field theory} and it may be a
relevant physical model at scales in between ${ \ell_{\rm P} }$ and ${
  \ell_{\rm LHC} }$. In fact, one of the main threads of research in
this field has been related to studies of energetic cosmic rays, as we
will discuss further below. In
the following we will study this relationship in some detail. These
field theories provide fruitful avenues of exploration for several
reasons, that will be explained in more depth below. Firstly, some
quantum field theories are better behaved on noncommutative spacetime
than on ordinary spacetime. In fact, some are completely finite, even
non-perturbatively. In this manner spacetime noncommutativity presents
an alternative to supersymmetry or string theory. Secondly, it is a
useful arena for studying physics beyond the standard model, and also
for standard physics in strong external fields. Thirdly, it sheds
light on alternative lines of attack to addressing various
foundational issues in quantum field theory, for instance the
renormalization and axiomatic programmes. Finally, it naturally
relates field theory to gravity. Since the field theory may be easier
to quantize, this may provide significant insights into the problem of
quantizing gravity.

This survey is intended for high-energy physicists who are
non-specialists in noncommutative field theory. As such, technical
details are kept to a minimum and we refer to various other literature
throughout for the relevant formalisms. Foundational aspects of
noncommutative quantum field theory are treated
in~\cite{Douglas01,Szabo03}. Brief reviews of the relationship between
spacetime noncommutativity and quantum gravity may be found
in~\cite{Szabo06,MullerHoissen08}. A detailed overview of the
renormalization programme for noncommutative field theories is given
in~\cite{Rivasseau07}. The connection between spacetime
noncommutativity and experimental signatures of Lorentz violation are
treated in~\cite{Bietenholz08}. The restoration of Lorentz symmetry is
related to the problem of renormalization of noncommutative field
theories in~\cite{Szabo07}. The interplay between noncommutative field
theory and physics in strong magnetic fields is discussed
in~\cite{Szabo04}. There are many other topics in this vast field that
are not touched upon in this review. Furthermore, the bibliography is
not meant to be exhaustive, and we apologise in advance to those
concerned for the omissions.

\bigskip

\section{Spacetime quantization\label{Spacetimequant}}

\noindent
In this section we will provide a (non-exhaustive) list of contexts in
which noncommutative spacetimes play a prominent role, particularly
those which can be argued to arise in various approaches to quantum
gravity, as well as the applications mentioned in~\S\ref{Intro}. The
aim is to set up some explicit models where noncommutative geometry is
naturally interlaced with the problem of quantizing gravity, at least
at a kinematical level. Later on we shall examine this relationship
from a more dynamical perspective.

\subsection{Snyder's spacetime\label{Snyder}}

The idea of spacetime noncommutativity is in fact very old. It is
usually attributed to Heisenberg who proposed it in the late 1930's as
a means of regulating the ultraviolet divergences which plague quantum
field theory. Heisenberg suggested this idea in a letter to Peierls,
who actually applied it in a non-relativistic context of electronic
systems in external magnetic fields (we will come back to this
application later on). Peierls told Pauli about the idea, who then
told Oppenheimer. Oppenheimer gave the problem to his graduate student
Snyder, whose is acredited today with the original paper on
noncommutative spacetime~\cite{Snyder47}. 

In a suitable basis, the algebra underlying Snyder's spacetime may be
presented as a modification of the phase space canonical commutation
relations given by 
\begin{eqnarray}
\big[x^\mu\,,\,x^\nu\big]&=&\ii{\ell^2\,\hbar^{-1}}\,
\big(x^\mu\,p^\nu-x^\nu\,p^\mu\big) \ , \nonumber \\[4pt]
\big[x^\mu\,,\,p_\nu\big]&=&\ii\hbar\,\delta^\mu{}_\nu+
\ii{\ell^2\,\hbar^{-1}}\,p^\mu\,p_\nu \ , \nonumber \\[4pt]
\big[p_\mu\,,\,p_\nu\big]&=&0 \ .
\label{Snyderalg}\end{eqnarray}
This algebra involves a fundamental minimal length ${\ell}$, the scale
of noncommutativity as in \S\ref{Intro}, such that the ``classical''
phase space of quantum mechanics is recovered at $\ell=0$. The
original motivation behind these relations was that the introduction
of the length scale $\ell$ is tantamount to regarding hadrons in
quantum field theory as extended objects, because at the time
renormalization theory was regarded as ``a distasteful
procedure''~\cite{Snyder47}. The commutation relations
(\ref{Snyderalg}) describe a discrete spacetime which is compatible
with Lorentz invariance. 

There is a natural way to derive the algebra (\ref{Snyderalg}) in
$3+1$-dimensions in terms of a dimensional reduction from
$4+1$-dimensions as the angular momentum generators of the
higher-dimensional Lorentz group $SO(1,4)$~\cite{Yang47} (which
clarifies at least the first relation in (\ref{Snyderalg})). In this
way the generators of Snyder's spacetime are naturally interpreted as
the generators which preserve a four-dimensional de~Sitter space
inside five-dimensional Minkowski space, and are naturally invariant
under both five-dimensional and four-dimensional Lorentz
transformations. This reduction is consistent with the fact that the
phase space in this model contains a curved momentum space, as
signified by the momentum dependence on the right-hand side of the
commutation relations (\ref{Snyderalg}).

The death of Snyder's spacetime, within the context in which it was
proposed at the time, was triggered by the eventual success of the
renormalization programme in quantum field theory. The historic
overwhelmingly successful agreement with experiment of the
calculations of both the Lamb shift in hydrogen and the anomalous
magnetic moment of the electron from radiative corrections in quantum
electrodynamics put the more conventional regularization techniques at
the forefront of modern quantum field theory. However, this spacetime
emerged again more than half a century later in the somewhat
unexpected and surprising context of quantum gravity. 

\subsection{$\kappa$-Minkowski spacetime\label{kappaMink}}

The phase space commutation relations of the $\kappa$-Minkowski
spacetime can be written in the bicrossproduct basis
as~\cite{Lukierski91,Majid94} 
\begin{eqnarray}
\big[x^\mu\,,\,x^\nu\big]&=&\frac\ii{\kappa}\,\big(x^\mu\,\xi^\nu-
x^\nu\,\xi^\mu\big) \ , \nonumber \\[4pt]
\big[x^\mu\,,\,p_\nu\big]&=&\ii\hbar\,\delta^\mu{}_\nu+
\frac\ii{{\kappa}}\,
\big(p^\mu\,\xi_\nu+p_\nu\,\xi^\mu\big) \ , \nonumber \\[4pt]
\big[p_\mu\,,\,p_\nu\big]&=&0 \ ,
\label{kappaMinkalg}\end{eqnarray}
where the lightlike four-vector $\xi^\mu$ has components $\xi^0=1$,
$\xi^i=0$, and $\kappa$ is a mass scale. In the notation of
\S\ref{Snyder}, we identify $\kappa\simeq\hbar/\ell$ as a very large
energy. In a somewhat less covariant formulation than in
(\ref{kappaMinkalg}), only commutators involving the time
components $x^0=c\,t$ and $p_0=E/c$ are deformed by the mass scale in
(\ref{kappaMinkalg}). In fact, this spacetime is equivalent to
Snyder's spacetime~\cite{KowalskiGlikman02} through a mapping between
the generators of the algebras (\ref{Snyderalg}) and
(\ref{kappaMinkalg}). This may seem surprising, given that the Snyder
coordinates are determined by Lorentz transformations preserving
four-dimensional de~Sitter space, rather than Minkowski space, but the
mapping is determined by highly non-local transformations involving,
for example, momentum-dependent redefinitions of the spacetime
coordinates. 

The $\kappa$-Minkowski algebra has been of interest lately as the
noncommutative spacetime underlying certain realizations of doubly
special relativity~\cite{AmelinoCamelia01}. These modifications of
special relativity contain, in addition to the speed of light $c$, a
second, short-distance observer-independent length scale, usually
identified as the Planck scale $\ell=\ell_{\rm P}$. They have been
proposed as a consistent description of quantum gravity in flat
spacetime, which are thereby amenable to phenomenological tests. In
these models the effective spacetime metric, or alternatively the
Poincar\'e symmetry of flat Minkowski spacetime, is deformed by energy
according to the relations (\ref{kappaMinkalg}), reflecting again the
curvature of momentum space. See~\cite{Ghosh06,Ghosh07} for an
explicit canonical framework which reflects the deformed symmetries of
doubly special relativity as a non-linear realization of the Lorentz
group.

This noncommutative deformation of Minkowski space modifies the
dispersion relations to~\cite{Magueijo02} 
\beq
E^2\=\mbf p^2\,c^2-m^2\,c^4\,\Big(1-\frac{\xi\cdot p}
{{\hbar\,\kappa}}\Big)^2 \ .
\label{kappaMinkdisp}\eeq
Thus the speed of a photon in $\kappa$-Minkowski spacetime depends on
its energy {$E$}. This is consistent because the dependence is through
the small dimensionless combination {$\ell_{\rm P}\,E$}. These
predictions have been tested with some success against measurements of
astrophysical gamma-ray bursts as measured, for example, by the GLAST
and MAGIC telescopes~\cite{Bietenholz08}. In the low-energy limit
$\kappa\to\infty$, whereby the algebra (\ref{kappaMinkalg}) reduces to
the usual phase space commutation relations of ordinary Minkowski
space, these results are all in agreement with the macroscopic
predictions of classical general relativity.

\subsection{Three-dimensional quantum gravity\label{3D}}

Perhaps the most precise dynamical realization of the connections
between noncommutative field theory and quantum gravity alluded to in
\S\ref{kappaMink} comes from a non-perturbative model of quantum
gravity in \emph{three} spacetime dimensions~\cite{Freidel05}. One
starts with a spin foam model of three-dimensional gravity coupled to
spinless matter, and then integrates out the gravitational degrees of
freedom in the path integral to produce an effective action for the
scalar fields. In this way one finds that the effective interactions
can be encoded in a scalar field theory on the $SO(1,2)$ Lie algebra
noncommutative spacetime
\begin{eqnarray}
\big[x^\mu\,,\,x^\nu\big]&=&\ii{\ell}\,\epsilon^{\mu\nu\lambda}\,
x_\lambda \ , \nonumber \\[4pt]
\big[x^\mu\,,\,p_\nu\big]&=&\ii\sqrt{\hbar^2-{\ell^2}
\,p^2}~\delta^\mu{}_\nu-
\ii{\ell}\,\epsilon^\mu{}_{\nu\lambda}\,p^\lambda \ , \nonumber \\[4pt]
\big[p^\mu\,,\,p^\nu\big]&=&0 \ .
\label{LiealgNC}\end{eqnarray}
The three-dimensional spacetime algebra (\ref{LiealgNC}) also arises
in the context of Lorentz covariant spacetime uncertainly
relations~\cite{Sasakura00} (such relations are discussed in more
detail below). This spacetime again has a $\kappa$-deformed Poincar\'e
symmetry, together with modified dispersion relations
\beq
E^2\=\mbf p^2\,c^2-\Big(\frac{\sinh({\ell\,\hbar^{-1}}\,m\,c^2)}
{{\ell\,\hbar^{-1}}}\Big)^2 \ .
\label{3DQGdisp}\eeq
This dispersion relation matches (\ref{kappaMinkdisp}) at leading
orders in the limit $\ell\to0$.

Thus in this case, doubly special relativity arises in the low-energy
limit of quantum gravity. The mechanism behind this emergence is an
effective noncommutative quantum field theory. The properties of this
noncommutative field theory and its precise relationship to
$2+1$-dimensional gravity is currently under active investigation. For
example, certain properties of the (curved) momentum space associated
to (\ref{LiealgNC}) indicate that some modifications seem
necessary. The periodicity of the momentum space ruins unitarity of
the noncommutative field theory, while the absence of arbitrary
negative energy means that it cannot correspond to the momentum space
of massive particles coupled to $2+1$-dimensional gravity. These
problems are analysed in~\cite{Sasai09}, where it is proposed that an
extension of the momentum space of the noncommutative field theory to
its universal covering group may resolve these problems.

\subsection{Spacetime uncertainty principle\label{Spacetimeuncert}}

Despite their appeal as direct relatives of quantum gravity and
high-energy astrophysical processes, the models of noncommutative
spacetime we have thus far considered are far too complex to be useful
for detailed direct study of the dynamics of quantum fields defined on
them. They are best handled using Hopf algebraic techniques and
methods of braided quantum field theory~\cite{Oeckl01,Sasai07}. We
seek a somewhat simpler setting which is amenable to the standard
perturbative methods of quantum field theory, but at the same time
still captures the nature of Planck scale physics. We will then be
able to study to what extent the field theoretic model captures the
dynamics of gravitational interactions. The models which we will
discuss in detail in the remainder of this survey can be motivated
within the above context by the simple semi-classical argument
of~\cite{Doplicher95}. This argument is based on combining only
fundamental postulates of general relativity with quantum mechanics,
and it demonstrates that spacetime quantization, and more generally
noncommutative geometry, is expected to be a generic feature of any
theory of quantum gravity, consistent with what we observed in
\S\ref{3D}.

Suppose that we try to probe physics at the Planck scale {$\ell_{\rm
    P}$}. Then the Compton wavelength of any probe must be smaller
than $\ell_{\rm P}$. But this means that there is a huge mass
${m\geq\hbar/\ell_{\rm P}\,c }$ concentrated in a tiny volume
${\ell_{\rm P}^3}$, and the energy density is large enough that it
forms a {\it black hole}. The event horizon has size determined by the
Schwarzschild radius which is of order $m$, and it thereby hides the
measurement we set out to make as no information can escape from the
interior of the black hole through its horizon. To avoid this problem,
there must be a mechanism in place which prevents the occurence of
such a gravitational collapse.

The mechanism proposed by~\cite{Doplicher95} is to introduce a
fundamental length $\ell$, that we may wish to identify with the
Planck scale $\ell=\ell_{\rm P}$, which limits both spatial and
temporal localization via the spacetime uncertainty relations
\beq
\Delta x^0\,\sum_i\,\Delta x^i\geq\ell_{\rm P}^2 \qquad \mbox{and}
\qquad 
\sum_{i<j}\,\Delta x^i\,\Delta x^j\geq\ell_{\rm P}^2 \ .
\label{uncertprop}\eeq
Then, if we wish to make a measurement along, say, the $j$-th
direction of space, the energy of our probe will spread out over a
perpendicularly directed disk of radius {$\Delta x^i$}, such that the
induced gravitational potential vanishes as we try to localize the
space coordinate $x^j$, i.e. in the limit where $\Delta
x^i\rightarrow\infty$. A similar reasoning applies to localization of
the time coordinate $x^0$. As indicated in
\S\ref{Intro}, a concrete model of quantum spacetime which fulfills
these requirements, and which is good for studying quantum field
theory thereon, has the quantum phase space commutation relations
\begin{eqnarray}
\big[x^\mu\,,\,x^\nu\big]&=&\ii\theta^{\mu\nu}\={\rm constant} \ ,
\nonumber \\[4pt]
\big[x^\mu\,,\,p_\nu\big]&=&\ii\hbar\,\delta^\mu{}_\nu \ ,
\nonumber\\[4pt]
\big[p_\mu\,,\,p_\nu\big]&=&0 \ ,
\label{DFRcommrels}\end{eqnarray}
with
\beq
\theta_{\mu\nu}\,\theta^{\mu\nu}\=0 \qquad \mbox{and} \qquad
\epsilon_{\mu\nu\lambda\rho}\,\theta^{\mu\nu}\,
  \theta^{\lambda\rho}\= -8\ell_{\rm P}^4 \ .
\label{thetaellP}\eeq
One of the main simplifications arising in this class of spacetimes is
that their momentum spaces are \emph{flat}, in contrast to those
considered earlier in this section. Only the commutators between two
spacetime coordinates are deformed. 

In the original treatment of~\cite{Doplicher95}, the constant
noncommutativity parameters $\theta^{\mu\nu}$ are treated as
``dynamical'' variables, in the sense that their constant values are
allowed to vary subject to the constraints (\ref{thetaellP}). This
means that the algebra of spacetime coordinates 
$x^\mu$ is enlarged to include $\theta^{\mu\nu}$ on an equal footing
as central elements, so that
\beq
\big[x^\mu\,,\,\theta^{\nu\lambda}\big]\=0 \ .
\eeq
In this enlarged algebra, the action of a Lorentz transformation on
$x^\mu$ can be compensated in the commutation relations
(\ref{DFRcommrels}) by letting $\theta^{\mu\nu}$ transform as a
covariant tensor of rank two under $SO(1,3)$. In this way, the quantum
spacetime is compatible with Lorentz invariance. However, in the
following we will instead think of $\theta^{\mu\nu}$ as a \emph{fixed}
and \emph{arbitrary} background tensor field, at the cost of breaking
Lorentz symmetry and the explicit uncertainty relations
(\ref{uncertprop}). Such a Lorentz violation could have experimental
signatures, as we will discuss in \S\ref{NC-signs}. The motivation
behind this restriction comes from various physical contexts in which
systems are subjected to strong external background fields, some
examples of which are discussed below. It also simplifies somewhat the
analysis of the resulting noncommutative quantum field theories.

\subsection{Physics in strong magnetic fields\label{Landau}}

The simplest prototypical example whose dynamics induces a quantum
space is the \emph{Landau problem}. The Landau problem considers the
quantum mechanics of a non-relativistic system of electrons
$$
\mbox{\includegraphics[width=2.5in]{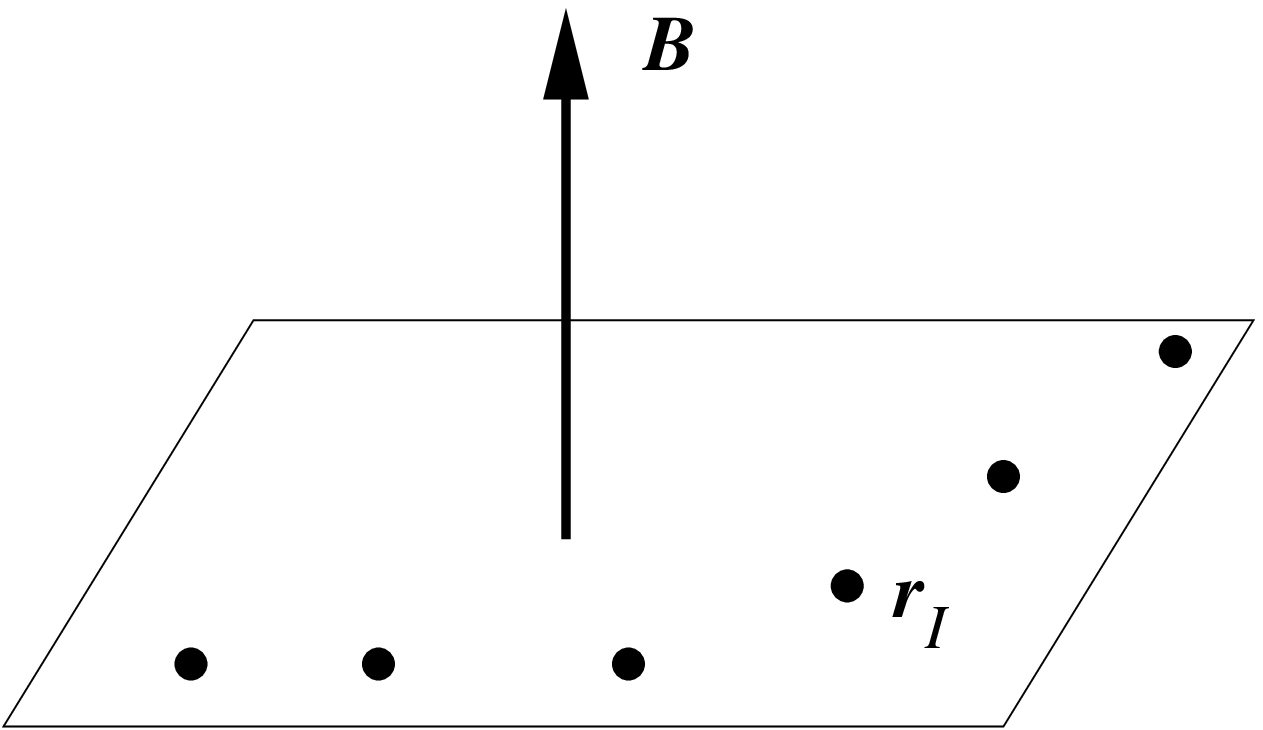}}
$$
confined to a plane $\mbf x=(x,y,0)$ and subjected to a uniform
background magnetic field of strength $B$ which is applied in the
direction perpendicular to the plane of motion. The lagrangian for
each electron is given by 
\beq
\mathcal{L}_m\=\frac m2\,\dot{\mbf x}{}^2-\frac ec\,\dot{\mbf
  x}\cdot\mbf A \ ,
\label{LandauLm}\eeq
where 
\beq
A_x\=-\frac B2\, y \qquad \mbox{and} \qquad A_y\=\frac B2\,x
\eeq
are the non-zero components of the corresponding vector potential in
the symmetric gauge.

The quantization of this system is an elementary exercise in senior
undergraduate level quantum mechanics. One can map the Hamiltonian of
this model onto that of a harmonic oscillator, whose spectrum yields
the Landau levels. The lowest Landau level is the ground state of this
harmonic oscillator. In the strong field limit {$e\,B\gg m$}, the
spacing between Landau levels becomes infinite and the spectrum
projects onto the lowest Landau level. The lagrangian (\ref{LandauLm})
in this regime truncates to
\beq
\mathcal{L}_0\=-\frac{e\,B}{2c}\,(\dot x\,y-\dot y\,x) \ .
\eeq
This reduced Lagrangian is of first order in time derivatives. The
phase space therefore becomes degenerate and collapses onto the
configuration space. Thus canonical quantization gives a {\it
  noncommutative space} having the commutators
\beq
[x,y]\=\ii\theta \qquad \mbox{with} \quad
\theta\=\frac{\hbar\,c}{e\,B} \ .
\label{Landaucommrels}\eeq

This observation can have many applications to systems which are
subjected to strong magnetic fields. As mentioned in \S\ref{Intro},
the first application was the famous Peierls
substitution~\cite{Peierls33}. This utilizes the Schr\"odinger
representation of the Heisenberg commutation relations
(\ref{Landaucommrels}) to compute the first order energy shift due to
an impurity potential {$V(x,y)$} in perturbation theory of the lowest
Landau level of the electronic system. More recently, based on the
analogy between the canonical quantization of the Landau problem above
and that of Chern--Simons gauge theory~\cite{Dunne89}, it has been used
to motivate a noncommutative Chern--Simons theory formulation of the
fractional quantum Hall effect which has been shown to provide a much
better microscopic description of the dynamics than the commutative
effective field theory~\cite{Susskind01}--\cite{Hellerman01}. It may
also be useful for studying certain models of charged polymer growth
in magnetic fields in terms of random walks with self-avoiding
interaction~\cite{Duplantier06}, and it may even find applications to
the description of bound states in non-abelian gauge theories, in
particular to quark
confinement~\cite{Gorbar05}. See~\cite{Rivasseau07,Szabo04} for
further discussion of these applications.

\subsection{Noncommutative geometry in string theory\label{NCstring}}

Our final example is the context in which noncommutative spaces arise
in string theory, which has been mostly responsible for the huge surge
of activity in noncommutative field theory over the past
decade. Strings naturally come with a length scale $\ell=\ell_s$,
their intrinsic length
$$
\mbox{\includegraphics[width=2.5in]{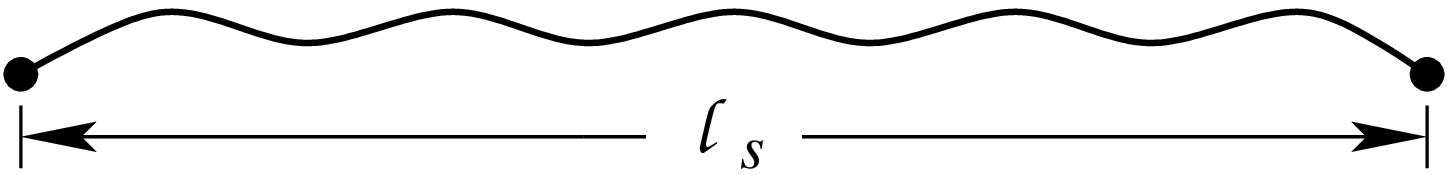}}
$$
whose square is inversely proportional to the string tension and such
that strings reduce to point particles in the limit
$\ell_s\to0$. Together with the string coupling constant, the string
length $\ell_s$ dynamically determines the Planck scale of the target
spacetime in which the strings live. 

It was realized from early analyses of very high-energy string
scattering amplitudes~\cite{Gross88,Amati89} that, in order to prevent
gravitational collapse, the objects in spacetime seen by strings must
obey a modification of the Heisenberg uncertainty relation given by 
\beq
\Delta x\geq\frac\hbar2\,\frac1{\Delta p}+\frac{\ell_s^2}{2\hbar}
\,\Delta p \ .
\label{modHeisrel}\eeq
At low energies this agrees with the standard phase space
quantization, but for energies $E\gg\hbar\,c/\ell_s$ it implies that
the extent of an observed object grows linearly with its
momentum. Varying the right-hand side of the uncertainty relation
(\ref{modHeisrel}) shows that it is minimized at $\Delta
p=\hbar/\ell_s$, and substituting this back into (\ref{modHeisrel})
shows that the minimum spatial resolution seen by a string probe is 
\beq
(\Delta x)_{\rm min}\=\ell_s \ .
\label{Deltaxmin}\eeq
This simply means that strings cannot probe distances below their
intrinsic size, and this fact was used to suggest very early on that
the concept of spacetime changes its meaning below the Planck scale,
due to the non-locality of string interactions. 

Via some basic kinematical conformal invariance arguments, it was
subsequently realized that in fact the target space probed by strings
must also be subjected to spacetime uncertainty
relations~\cite{Yoneya00}
\beq
\Delta x\,\Delta t\geq \frac{\ell_{\rm P}^2}c \ ,
\label{stringDeltaxDeltat}\eeq
analogous to what was observed in \S\ref{Spacetimeuncert}. Relations
such as these can be derived dynamically by looking at
non-perturbative degrees of freedom in string theory. This was done by
noting that sub-Planckian scales in string theory are probed by {\it
  D-branes}~\cite{Douglas97}, whose dynamics thus enable one to
provide a microscopic derivation of these and other modified
uncertainty relations~\cite{Li96,Mavromatos98}. A D-brane is a
boundary condition for open strings. We can regard D-branes as
hypersurfaces in spacetime
$$
\mbox{\includegraphics[width=2.5in]{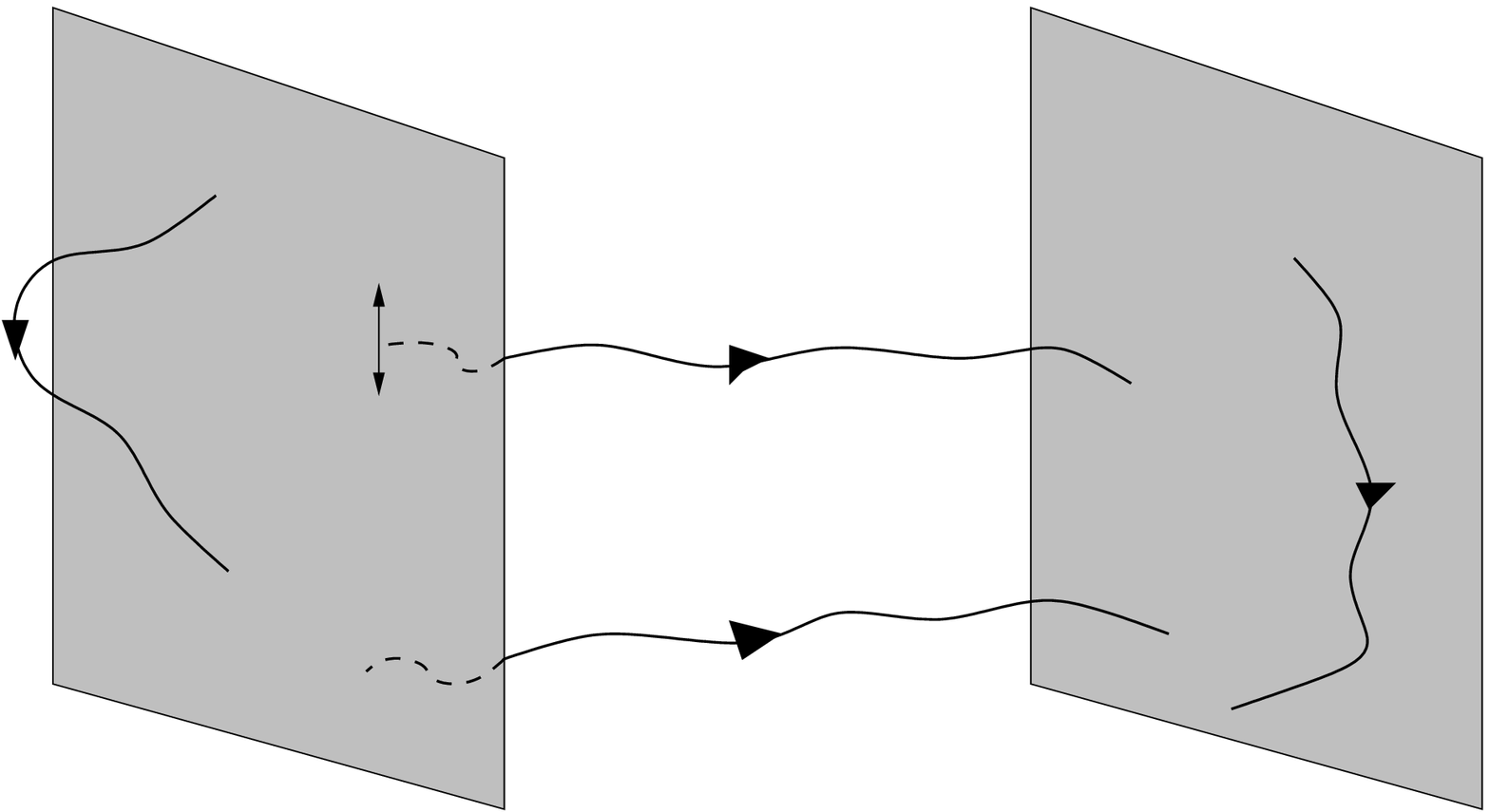}}
$$
onto which the endpoints of open strings attach. The quantum dynamics
of open strings induce fields which live on the branes. These include,
among many others, Chan--Paton gauge fields $A_\mu$, scalar fields
$X^m$ describing the location of the D-brane in spacetime, and also
fermion fields $\psi$ in the supersymmetric case. In the low-energy
limit, the spacetime dynamics of the open strings are described by an
effective field theory living on the D-branes, which dynamically
describes how the shape and size of the D-brane fluctuates in
spacetime. In particular, by considering string theory with D-branes
and certain background ``magnetic'' fields $B$ induced by the closed
string sector, the low-energy limit is described by a \emph{field
  theory on a noncommutative space}~\cite{Schomerus99,Seiberg99}, via
a mechanism analogous (though somewhat more involved) to that
described in \S\ref{Landau}. It is this D-brane field theory that we
will be interested in for the remainder of this survey. 

The derivation just sketched is what led to the huge intensity of
research starting at the end of the last millenium, and has today
developed into its own branch of physics called ``noncommutative field
theory''. It gives an explicit realization of the manner in which
worldvolume field theories are altered by quantum gravitational
effects. Since the closed string sector induces the geometric fields
of the target spacetime, D-branes can interact gravitationally with
one another by exchanging closed strings. This couples the D-brane
field theory to gravity, and will be one of the ways in which field
theories on quantized spaces are intimately connected to gravity. 

\bigskip

\section{Field theory on quantized spacetimes\label{NC-QFT}}

\noindent
In this section we will describe some aspects of the perturbative
dynamics of the noncommutative quantum field theories we derived in
the previous section. Later on we will see how these novel features
can be reinterpreted in the context of quantizing gravity directly
from noncommutative field theory, and also in terms of deformed
dispersion relations such as (\ref{kappaMinkdisp}) and
(\ref{3DQGdisp}). In this section we will only consider scalar fields,
defering the discussion of gauge fields to the next
section. Henceforth, unless otherwise indicated, we shall work in
euclidean signature and with natural units in which $\hbar=c=1$. We
will always assume that the antisymmetric matrix $\theta^{\mu\nu}$ is
invertible.

\subsection{Formalism}

Let us begin by describing the basic aspects of formulating
noncommutative quantum field theory, as a deformation of ordinary
field theory (see~\cite{Douglas01,Szabo03} for further details). The
noncommutative deformation is implemented by replacing the usual
pointwise product of a pair of fields $\phi$ and $\psi$ by the
``star-product'' 
\beq
\phi(x)\,\psi(x)~\longrightarrow~(\phi\star\psi)(x)\=\phi(x)\,
\psi(x)+O\big(\theta\,,\,\partial\phi\,\partial\psi\big) \ .
\label{starprod}\eeq
The star-product defines an associative but noncommutative
multiplication on the space of fields, which deforms the usual product
that is recovered at $\theta=0$. We will regard it as containing
infinitely-many bi-derivative terms. There are several ways to express
the star-product, depending on the domain of fields used. For most of
our considerations below, we will use an integral representation of
the star-product, which makes it a non-local operation. The two
representations coincide on analytic fields. Here we shall only need
its simplest incarnation. Denoting by $\tilde\phi$ the Fourier
transform of the field $\phi$, the momentum space version of
(\ref{starprod}) modifies the Fourier convolution product as 
\beq
\tilde\phi(k)\,\tilde\psi(q)~\longrightarrow~
\tilde\phi(k)\,\tilde\psi(q)~
\e^{\ii k\times q} \qquad \mbox{with} \quad k\times
q\=\mbox{$\frac12$}\, k_\mu\,\theta^{\mu\nu}\,q_\nu \ .
\label{starconvprod}\eeq

An easy calculation shows that this definition gives the required
commutation relations 
\beq
\big[x^\mu\,,\,x^\nu\big]_\star\=x^\mu\star x^\nu-x^\nu\star
x^\mu\=\ii\theta^{\mu\nu} \ .
\label{starcomm}\eeq
Given any ordinary field theory with action $S$, one obtains a
noncommutative field theory simply by replacing all products of fields
occuring in $S$ with star-products. This deformation only affects the
interactions of fields, owing to the identity 
\beq
\int\,\phi\star\psi\=\int\,\phi\,\psi
\label{intphipsi}\eeq
which is straightforwardly derived via integration by parts. This
property is particular to the relatively simple noncommutative spaces
we have restricted to, and it is another reflection of the flatness of
momentum space (equivalently, the invariant line element ${\rm d}s^2$
is undeformed). It implies that the underlying free field
theory is unaltered by the effects of noncommutativity. This is one
reason why the quantum dynamics of these noncommutative field theories
are tractable, at least in perturbation theory.

\subsection{UV/IR mixing\label{UVIRmixing}}

Let us now describe the most notorious problem which arose in some of
the first studies of perturbative noncommutative field
theory. Consider, for example, the simplest instance of a real scalar
$\lambda\,\phi^{\star n}$ theory. The interaction vertices are given
by the Feynman rules
\unitlength=1.00mm
\linethickness{0.4pt}
\beq
\begin{picture}(100.00,25.00)
\thinlines
\put(32.00,12.00){\line(-1,1){10.00}}
\put(32.00,12.00){\line(-1,-1){10.00}}
\put(32.00,12.00){\circle*{1.50}}
\put(18.00,22.00){\makebox(0,0)[l]{{$k_1$}}}
\put(18.00,2.00){\makebox(0,0)[l]{{$k_2$}}}
\put(26.00,12.00){\makebox(0,0)[l]{{$\lambda$}}}
\put(32.00,12.00){\line(1,1){10.00}}
\put(32.00,12.00){\line(1,-1){10.00}}
\put(40.00,12.00){\makebox(0,0)[l]{$\vdots$}}
\put(44.00,2.00){\makebox(0,0)[l]{{$k_3$}}}
\put(44.00,22.00){\makebox(0,0)[l]{{$k_n$}}}
\put(48.00,12.00){\makebox(0,0)[l]{{$~=~\lambda~
\e^{\ii\sum\limits_{I<J}k_I\times k_J} \ , $}}}
\end{picture}
\label{phinvertex}\eeq
with the usual momentum conservation {$k_1+k_2+\ldots+k_n=0$} since
the noncommutative space (and the noncommutative field theory) is
translationally invariant. The phase factors in these interactions
become effective at energies {$E$} with {$E\,\sqrt{\theta}\gg1$}. 

In contrast to the commutative case, the vertex (\ref{phinvertex})
depends on the ordering of the momenta $k_I$. However, it depends only
on their \emph{cyclic} ordering. In analogy with how one handles the
perturbative dynamics of multi-colour field theories, it is convenient
to represent each Feynman diagram by fattening its lines and making
the diagram into a ribbon graph. Then the set of all graphs comes with
a notion of planarity. \emph{Planar graphs} are those which can be
drawn on the plane without crossing lines. They coincide with the
ordinary graphs at $\theta=0$, times possible phase factors depending
only on the momenta of the external
legs~\cite{Filk96,Ishibashi99}. This shows that noncommutativity does
not tame the ultraviolet divergences of quantum field theory as
originally hoped, at least for this simple class of noncommutative
spaces, and the noncommutative field theory contains at least the same
divergences. 

What is even worse is the behaviour of the \emph{non-planar graphs},
those which cannot be drawn on the plane without crossing lines. In
these diagrams, the phase factors of (\ref{phinvertex}) depend on the
virtual momenta of internal loops. Although these phase factors
succeed in damping out the high-energy behaviour of the graphs through
rapid oscillations and would seem to provide a natural ultraviolet
cutoff $1/\theta\cdot k$, they become ineffective for vanishing
momenta and the original ultraviolet divergences reappear as infrared
divergences. This phenomenon is called UV/IR
mixing~\cite{Minwalla99}. It leads to severe problems when one-loop
non-planar diagrams occur as subgraphs in two-loop and higher order
graphs, giving uncontrollable divergences which prevent the
renormalization of higher orders in perturbation
theory~\cite{VanRaamsdonk00}. In other words, the introduction of an
ultraviolet cutoff $\Lambda$ in the field theory induces an effective
infrared cutoff 
\beq
\Lambda_0\=\frac1{\theta\,\Lambda} \ .
\label{Lambda0IR}\eeq
This mixing prohibits the use of standard renormalization schemes,
such as the wilsonian approach, which require a clear separation of
energy scales. It follows that \emph{the field theory cannot be
  renormalized}. This problem plagued noncommutative field theory for
many years. 

In order to understand how to overcome this serious problem, it is
useful to briefly look at the physics underlying the UV/IR mixing
phenomenon. The mixing is due to the inherent non-locality of the
quantum field theory. If $\phi$ and $\psi$ are fields which are both
supported in a region of small size $\Delta\ll\sqrt{\theta}$, then
their star-product {$\phi\star\psi$} is non-zero in a large region of
size {$\theta/\Delta$}. A drastic example of this occurs when both
fields are localized at a point, as then their interaction is
supported on all of space due to the identity 
\beq
\delta(x)\star\delta(x)\=\frac1{\det(\pi\,\theta)} \ .
\label{deltastarprod}\eeq
This means that the quanta governing the interactions of
noncommutative field theory in the ultraviolet must also be non-local
extended objects which are very different from the pointlike quanta
encountered in ordinary quantum field theory.

The nature of these new degrees of freedom can be understood via an
elementary calculation using (\ref{starcomm}) and the
Baker--Campbell--Hausdorff formula which gives the identity 
\beq
\e^{\ii k\cdot x}\star\phi(x)\star\e^{-\ii k\cdot
    x}\=\phi(x^\mu-\theta^{\mu\nu}\,k_\nu) \ .
\label{planewavetransl}\eeq
We interpret this to mean that the ultraviolet dynamics, in the regime
$E\gg\theta^{-1/2}$, of the fields behave as if they were supported on
extended, rigid rods whose size is proportional to their
momentum. These are ``dipoles'' with dipole moment
\beq
\Delta x^\mu\=\theta^{\mu\nu}\,k_\nu \ ,
\label{dipolemoment}\eeq
and they interact by joining at their
ends~\cite{SheikhJabbari99,Bigatti99}. This is analogous to the
electron-hole bound states in a strong magnetic field, as in the
systems described in \S\ref{Landau}, and it also agrees with the
modified uncertainty relation (\ref{modHeisrel}) derived from string
theory. Thus in the ultraviolet the natural degrees of freedom are
dipoles. On the other hand, the infrared dynamics, in the regime
$E\ll\theta^{-1/2}$ where the effects of noncommutativity are
negligible, is mediated by the elementary quantum fields $\phi$
themselves, with pointlike momenta $k_\mu$. This structure is in fact
suggested directly by the commutation relations (\ref{DFRcommrels}),
which lead to a noncommutative spacetime at high energies and an
almost commutative spacetime in the infrared regime. Although
completely different in their respective characteristics, it has been
suggested that there is a UV/IR ``duality'' relating the dynamics in
the two regimes~\cite{Rey02}. Evidence for this duality is provided
via perturbative calculations using explicit operators 
\beq
W_k[\phi]\=\int\,\exp\big(\ii|k|\,\phi(x)\big)
\label{Wkphi}\eeq
which create the noncommutative dipole degrees of freedom.

This simple physical picture suggests that the origin of the UV/IR
mixing problem is associated with the asymmetry between extended and
pointlike degrees of freedom in the different regimes of the quantum
field theory. One thereby seeks a covariant version of the field
theory which renders the ultraviolet and infrared regimes
indistinguishable~\cite{Langmann02}, and hence makes the UV/IR
``duality'' above into a true symmetry of the model. This
covariantization turns the infrared degrees of freedom into
extended objects by replacing their pointlike momenta with the
``Landau'' momenta 
\beq
k_\mu~\longmapsto~K_\mu\=k_\mu+B_{\mu\nu}\,x^\nu \ ,
\label{Landaumomenta}\eeq
analogously to the way in which the ultraviolet degrees of freedom extend
into dipoles. The additional non-degenerate constant antisymmetric
matrix $B_{\mu\nu}$ is in general independent of $\theta^{\mu\nu}$ and
can be interpreted as a ``magnetic'' background. Assuming that
$(x^\mu,k_\nu)$ form a canonical pair, the new momenta $K_\mu$
generate a ``noncommutative momentum space'' with the commutation
relations 
\beq
[K_\mu,K_\nu]\=2\ii B_{\mu\nu} \ ,
\label{NCmomsp}\eeq
in exactly the same in which noncommutative space arose in the Landau
problem of \S\ref{Landau}~\cite{Szabo04}. The covariant version of a
noncommutative complex scalar field theory thus consists in replacing
the hamiltonian of the usual kinetic term with the generalized Landau
hamiltonian defined by the square of gauge covariant derivatives in
the magnetic background according to (\ref{Landaumomenta}).

The real version of the duality covariant model is known as the
\emph{Grosse--Wulkenhaar model}~\cite{Grosse04}. Drawing on the
mapping between the Landau hamiltonian and the harmonic oscillator
hamiltonian, one considers a real $\lambda\,\phi_{2d}^{\star4}$-theory
in a background harmonic oscillator potential, defined by modifying
the scalar field kinetic term via the replacement 
\beq
\partial_\mu^2~\longmapsto~\partial_\mu^2+\frac{\omega^2}2\,\tilde
x_\mu^2 \qquad \mbox{with} \quad
\tilde x_\mu\=2\theta_{\mu\nu}^{-1}\,x^\nu \ .
\label{covho}\eeq
One then shows~\cite{Langmann02} that the full quantum field theory is
symmetric under symplectic Fourier transformation of the fields which
interchanges 
\beq
k_\mu~\leftrightarrow~\tilde x_\mu \ .
\label{symplinterchange}\eeq
Modifying the noncommutative field theory in this way gives the new
free propagators the correct decay behaviour, due to the confining
nature of the harmonic oscillator potential, that enables one to apply
wilsonian renormalization in a suitable duality-invariant basis of
fields. Note that while any confining potential could in principle be
used instead in (\ref{covho}) to give an effective infrared cutoff,
the harmonic oscillator potential is singled out as the unique one
which induces the UV/IR duality. 

\subsection{Renormalization\label{Renormalization}}

Let us now discuss some consequences of the UV/IR duality. The main
success of the duality covariant model is that \emph{it is
  renormalizable to all orders} in the $\phi^{\star4}$ coupling
constant $\lambda$~\cite{Grosse04}. The proof is obtained by an exact
mapping of the field theory onto a (infinite-dimensional) matrix
model, which naturally arises due to the infinite degeneracy of Landau
levels that provides a two-index basis set of complete Landau
wavefunctions for the expansion of the noncommutative fields. The
duality-invariant cutoff is then naturally taken to be the matrix size
$N$, and the wilsonian approach can be applied to the truncated model
as $N$ varies. This matrix model possesses several interesting
characteristics~\cite{Langmann04}. For example, the $N\times N$ matrix
model is related to an integrable KP-hierarchy and, for certain
choices of parameters, it is exactly solvable in the limit
$N\to\infty$ as the cutoff is removed. This remarkable reformulation
of noncommutative field theory without any reference to a background
spacetime has no analog in ordinary quantum field theory. It is the
crux of many important investigations and will arise again in the next
section.

The point $\omega=1$ in parameter space is special and is called the
self-dual point, because there the quantum field theory is completely
invariant under the UV/IR duality (without any rescalings). This point
corresponds to $\theta=2B^{-1}$ where the noncommutativity is induced
by the magnetic background of \S\ref{UVIRmixing}, as it happens in the
Landau problem (see (\ref{Landaucommrels})). At this point, a
remarkable property is found~\cite{Grosse05,Disertori06}. The
beta-functions for both the coupling constant $\lambda$ and the
oscillator parameter $\omega$ vanish to all orders of perturbation
theory. This implies that the renormalized coupling flows to a finite
bare coupling. The novel mechanism at work is that the wavefunction
renormalization compensates exactly the coupling constant
renormalization, such that the interaction $\lambda\,\phi^{\star4}$ is
invariant. In addition, the renormalized oscillator parameter flows to
the value $\omega=1$.

This boundedness of the renormalization group flow has the following
remarkable significance. To understand its implications, we need only
recall that a particle in the flow of a quadratic vector field goes to
infinity in a finite time. A similar feature was discovered in the
early days of quantum electrodynamics, where it was observed that the
coupling constant flows to infinity at very high, but finite,
energies. Thus the high-energy perturbation series does not make
sense, due to spurious poles that appear in a partial resummation of
the expansion which is possible in this energy regime. The
non-perturbative degrees of freedom corresponding to these poles is
called the Landau ghost or renormalons. The Landau ghost also plagues
ordinary $\phi^4$-theory in four dimensions, and it hinders a
non-perturbative definition of the quantum field theory. Were it not
for the discovery of asymptotic freedom in quantum chromodynamics, the
Landau ghost might have disseminated quantum field theory. What is
remarkable is that also in the duality covariant noncommutative field
theory, there are no renormalons. But here the field theory is not
asymptotically free, rather it is asymptotically safe. 

For these reasons, it is strongly believed that a non-perturbative
completion of these noncommutative quantum field theories is possible,
and intense research on this matter has been pursued over the past few
years. See~\cite{Rivasseau07} and references therein for a discussion
of these matters. From these perspectives, the duality covariant
models may teach important lessons for the consistent treatment of
ordinary quantum field theories. This programme would require a proper
understanding of the meaning of the parameters $\theta$ and $\omega$
at observable energies, and also a thorough understanding of how to
pass to our physical $3+1$-dimensional spacetime. The analytic
continuation of the duality covariant field theories to Minkowski
signature has been recently carried out in~\cite{Fischer09}, but the
renormalization of noncommutative quantum field theory in Minkowski
spacetime is still very much in its infancy. In particular, a simple
physical explanation of electric-type noncommutativity like the lowest
Landau level projection of \S\ref{Landau} is currently lacking.

UV/IR mixing is also known to occur on more complicated noncommutative
spaces, such as the $\kappa$-deformed space discussed in
\S\ref{kappaMink}~\cite{Grosse06}. The precise meaning of the duality
covariance in these and other instances is not very well understood,
and it may be related to some of the well-known position-momentum type
quantum group dualities. The UV/IR duality has been interpreted
recently in~\cite{Bieliavsky08} in terms of metaplectic
representations of the Heisenberg group, where the Grosse--Wulkenhaar
model on solvable symmetric spaces is formulated.

\bigskip

\section{Noncommutative gauge theory of gravity\label{NCGT-gravity}}
 
\noindent
Gauge theories on quantized spacetime are of particular interest for a
variety of reasons. They enable one to attempt to formulate more
realistic physical models which may provide a means for comparing with
measurable signatures of spacetime noncommutativity. This aspect will
be discussed in the next section. They are also the natural field
theories that arise on D-branes subjected to external
backgrounds~\cite{Seiberg99}, in the context discussed in
\S\ref{NCstring}. Given their role in string theory, it is natural to
expect that they may canonically couple to gravitational degrees of
freedom, and hence their quantization may teach us something about
quantum gravity. This is the theme which we shall have in mind in this
section. We will see that the UV/IR mixing phenomenon of the previous
section has a beautiful gravitational avatar in this context.

\subsection{Gauge interactions\label{gaugeint}}

The action for a $U(N)$ gauge field $A_\mu(x)$ in noncommutative
Yang--Mills theory is given by 
\beq
S\=-\frac1{4g^2}\,\int\,\Tr\,F_{\mu\nu}^2 \ ,
\label{SNCYM}\eeq
with the field strength tensor
\begin{eqnarray}
F_{\mu\nu}&=&\partial_\mu A_\nu-\partial_\nu
A_\mu-\ii[A_\mu,A_\nu]_\star \nonumber \\[4pt]
&=& \partial_\mu A_\nu-\partial_\nu
A_\mu-\ii[A_\mu,A_\nu] + O\big(\theta\,,\,(\partial A)^2\big) \ .
\label{NCfieldstrength}\end{eqnarray}
This model thus gives a modification of ordinary gauge theory by
infinitely many higher-derivative interaction terms. The gauge
invariance in this model is manifested by the invariance of
(\ref{SNCYM}) under the ``star-gauge transformations'' 
\beq
A_\mu~\longmapsto~U\star A_\mu\star U^{-1}+\ii U\star\partial_\mu
U^{-1} \qquad \mbox{with} \quad U\star U^\dag\=U^\dag\star U\=1 \ .
\label{stargaugetransf}\eeq
It follows that the gauge symmetry of noncommutative Yang--Mills
theory contains an intricate mixing of both colour and spacetime
transformations. It generates an infinite-dimensional unitary symmetry
group isomorphic to $U(\infty)$, as will become more apparent below
when we consider a certain matrix model formulation of this gauge
theory. Geometrically, star-gauge transformations generate certain
``deformed'' canonical transformations with respect to the Poisson
structure induced by the noncommutativity matrix
$\theta^{\mu\nu}$~\cite{Lizzi01}. 

Let us briefly summarize some of the novel features of noncommutative
gauge theories that distinguish them from their commutative
counterparts. Since the group $SU(N)$ does not close under the
star-product, because
$\det(\phi\star\psi)\neq\det(\phi)\star\det(\psi)$ in general, the
$U(1)$ sector cannot be decoupled from the $SU(N)$
sector~\cite{Armoni00}. The $U(1)$
coupling constant flows, and its beta-function is found to agree
\emph{precisely} with the beta-function of planar (large $N$) $SU(N)$
Yang--Mills theory~\cite{Hayakawa99} (the reason for this coincidence
will become evident below). Wilson loops have been observed to display
a phase structure~\cite{Alishahiha99,Bietenholz02}. They obey the
usual area law for small area loops, where the effects of
noncommutativity become negligible, while large area loops acquire an
imaginary Aharanov--Bohm phase with respect to the magnetic field
$B=1/\theta$ (again in agreement with (\ref{Landaucommrels})). UV/IR
mixing is present but occurs through a logarithmic
dependence~\cite{Matusis00} (in contrast to the power-law dependence
in scalar field theories) and only in the $U(1)$
sector~\cite{Armoni00}. This produces a strange infrared behaviour of
the ``photon'' in these gauge theories, which
may be compared to the high-energy behaviour of gamma-ray bursts
analogously to what we described in \S\ref{kappaMink}. We will return
to this point in the next section. The duality covariant version of
noncommutative gauge theory is not presently known. 

Let us also briefly mention a key tool for the analysis of
noncommutative Yang--Mills theory, the celebrated Seiberg--Witten
map~\cite{Seiberg99}. This transformation provides a one-to-one
correspondence between commutative and noncommutative gauge orbits of
gauge fields $A_\mu$, and is valid at length scales much longer than
the scale of noncommutativity $\sqrt\theta$. When applied to the
action (\ref{SNCYM}), it defines an ordinary gauge theory which is
``dual'' to noncommutative Yang--Mills theory. The dual gauge theory
is not a Yang--Mills theory, but contains many more higher order gauge
vertices. It has been utilized in a variety of contexts, and will play
a crucial role in our discussion later on. It has proved to be
particularly useful for producing ``phenomenological'' models which
could test the existence of spacetime noncommutativity in nature, for
example noncommutative extensions of the standard model which can
probe physics beyond the standard model~\cite{Calmet01}. We will also
return to this point in the next section.

\subsection{Gravity in noncommutative gauge theories}

It was realized early on that gravity is naturally contained in the
dynamics of noncommutative gauge theory, though the precise mechanism
was not initially clear. This is due to the spacetime transformation
properties of the gauge symmetry in these models. The key observation
is that spacetime translations of noncommutative gauge fields are
equivalent to gauge transformations. Using the identity
(\ref{planewavetransl}), the star-gauge transformation
(\ref{stargaugetransf}) by the star-unitary field 
\beq
U(x)\=\e^{\ii\theta_{\mu\nu}^{-1}\,a^\mu\,x^\nu}
\label{starunitarytransl}\eeq
is found to be
\beq
A_\mu(x)~\longmapsto~A_\mu(x+a)-\theta_{\mu\nu}^{-1}\,a^\nu \ .
\label{gaugetransl}\eeq
The constant shift in (\ref{gaugetransl}) drops out of the
noncommutative field strength tensor (\ref{NCfieldstrength}), and
hence the translational symmetry is a gauge
symmetry~\cite{Gross00}. This means that noncommutative gauge theories
provide ``toy models'' of general relativity, and may possibly enable
one to construct diffeomorphism invariant field theories.

The key to making this observation more precise is to be able to
promote the global translation symmetry to a local gauge symmetry, and
to extend the gauging to the full Poincar\'e symmetry of
four-dimensional spacetime. There have been various suggestions on how
to obtain Utiyama--Kibble type gauge theories of gravity along these
lines. For example, certain dimensional reductions of noncommutative
Yang--Mills theory from ten dimensions to four dimensions naturally
induce deformations of a Poincar\'e gauge theory of gravity, owing to
the occurence of teleparallism in noncommutative gauge
theory~\cite{Langmann01,Szabo06}. General relativity on noncommutative
spacetime has also been constructed by gauging the twist-deformed
Poincar\'e symmetry~\cite{Chaichian08}. Deformations of gravity can
moreover be induced from a noncommutative gauge theory with
position-dependent noncommutativity $\theta^{\mu\nu}(x)$ using the
Seiberg--Witten map~\cite{Marculescu08}. We will shortly see how an
analogous procedure can be used to extract general relativity more
precisely from the dynamics of noncommutative gauge
fields. Noncommutative gauge theories of gravity based on the
Lorentz group $SL(2,\C)$ with complex vierbein have been considered
in~\cite{Chamseddine03,Aschieri09}

\subsection{Spacetime disappears}

The crux of the identification of gravitational dynamics is the
background independent formulation of noncommutative gauge
theory~\cite{Seiberg99}, which allows a reformulation of the gauge
dynamics without any reference to a background spacetime. For this, we
introduce the ``covariant coordinates''~\cite{Madore00} 
\beq
X_\mu\=\theta_{\mu\nu}^{-1}\,x^\nu+A_\mu \ ,
\label{covcoords}\eeq
in terms of which the noncommutative field strength tensor
(\ref{NCfieldstrength}) can be written as 
\beq
F_{\mu\nu}\=-\ii[X_\mu,X_\nu]_\star+\theta_{\mu\nu}^{-1} \ .
\label{fieldstrengthcov}\eeq
This rewriting of the gauge fields again exploits the
translation-generating property (\ref{planewavetransl}) to write
derivatives as inner automorphisms of the algebra of fields. We may
thus reformulate noncommutative Yang--Mills theory with action
(\ref{SNCYM}) entirely using the operators $X_\mu$ which can be
regarded now as abstract objects of an infinite-dimensional matrix
algebra, without any reference to a spacetime dependence. 

The noncommutative gauge theory thereby becomes a \emph{matrix model}
with action 
\beq
S\=-\frac1{4g^2}\,\Tr\big(-\ii[X_\mu,X_\nu]+\theta_{\mu\nu}^{-1}
\big)^2 \ ,
\label{SIKKT}\eeq
where the infinite-dimensional trace implicitly hides the integration
over spacetime. This action actually arises in an independent context
as a twisted reduced model.  It is the dimensional reduction of
ordinary Yang--Mills theory to a point, i.e. with gauge fields which
do not depend on the spacetime coordinates, with the constant shift
$\theta_{\mu\nu}^{-1}$ the ``twist''. It is related to the IKKT matrix
model for the non-perturbative dynamics of Type~IIB
superstrings~\cite{Aoki99}. The equations of motion derived by varying
(\ref{SIKKT}) are 
\beq
[X_\mu,[X_\mu,X_\nu]]\=0 \ .
\label{eomIKKT}\eeq
In particular, the vacuum solution is given by matrices $X_\mu$ satisfying
\beq
[X_\mu,X_\nu]\=-\ii\theta_{\mu\nu}^{-1}
\label{Xvacuum}\eeq
which gives the absolute minimum value of the action
(\ref{SIKKT}). This solution is obtained from (\ref{covcoords}) with
$A_\mu=0$, and so in this context noncommutative spacetime arises as a
dynamical effect in the matrix model, analogously to the way it did in
\S\ref{3D}. The decomposition (\ref{covcoords}) then identifies the
gauge field degrees of freedom as fluctuations around the original
noncommutative spacetime coordinates $x^\mu$. 

The vacuum state is defined by a Heisenberg algebra (\ref{Xvacuum}),
which has only infinite-dimensional representions. To have a
constructive definition of noncommutative gauge theory, we would like
to define the natural regularization (as in \S\ref{Renormalization})
by cutting off the matrix rank at some finite value $N$. As
formulated, this is not possible because the vacuum equation
(\ref{Xvacuum}) has no finite rank $N\times N$ solutions about which
we can expand to uncover the dynamics of our original gauge theory. A
non-perturbative definition was proposed in~\cite{Ambjorn99} based on
the simple observation that while the Heisenberg commutation relations
(\ref{Xvacuum}) do not admit any finite-dimensional representations,
its ``exponentiated version'' (the Weyl algebra) does for certain
values of the noncommutativity parameter. Thus instead of the
hermitean degrees of freedom $X_\mu$, one considers a unitary matrix
model involving $N\times N$ unitary matrices $U_\mu$ with action 
\beq
S\=-\frac1{4g^2}\,\sum_{\mu\neq\nu}\,\e^{-2\pi\ii Q_{\mu\nu}/N}~
\Tr\big(U_\mu\,U_\nu\,U_\mu^\dag\,U_\nu^\dag\big) \ .
\label{STEK}\eeq
The matrix model (\ref{SIKKT}) is recovered by setting
\beq
U_\mu\=\e^{\ii a\,X_\mu}
\label{UXa}\eeq
and expanding to leading orders in the lattice spacing $a$. This
identifies the noncommutativity parameter as 
\beq
\theta^{-1}_{\mu\nu}\= \frac{2\pi\,Q_{\mu\nu}}{N\,a^2} \ .
\label{thetaQ}\eeq
The non-trivial vacuum state is now given in terms of the well-known
$SU(N)$ 't~Hooft twist-eating solutions ($N\times N$ clock and shift
matrices). 

The matrix model with action (\ref{STEK}) is known as the twisted
Eguchi--Kawai model. It can also be independently derived as the
one-plaquette reduction of Wilson's lattice gauge theory with
background 't~Hooft flux $Q_{\mu\nu}$ (the ``twist''). It was
originally proposed as a close relative of the planar $N\to\infty$
limit of multi-colour Yang--Mills theory, whose dynamics are more
tractable both analytically and numerically. In the present context,
this same matrix model admits a concrete \emph{finite} $N$
interpretation as a noncommutative version of lattice gauge theory,
with the matrix rank $N$ giving the finite size of the
lattice~\cite{Ambjorn99}. The automatic requirement of a (periodic)
lattice of \emph{finite} size for $N$ finite is a non-perturbative
manifestation of UV/IR mixing. The unitary matrix model has proven to
be useful for numerical studies of noncommutative gauge theory (see
e.g.~\cite{Bietenholz02a} and references therein). 

There are two distinct scaling limits of the matrix model (\ref{STEK})
with $N\to\infty$, $a\to0$, which explains the intimate relationship
between the noncommutative and planar Yang--Mills theories that we
noticed earlier. Taking the large $N$ limit first sends
$\theta\to\infty$, and gives the 't~Hooft limit of large~$N$
Yang--Mills theory (the limit $\theta\to\infty$ is also known to kill
all non-planar graphs of the noncommutative field
theory~\cite{Szabo03}). On the other hand, we can consider a special
double-scaling limit in which the combination $\sqrt{N}\,a^2$ is kept
finite. Then the noncommutativity $\theta$ remains finite and one
recovers noncommutative Yang--Mills theory. 

\subsection{Gravitational dynamics and UV/IR mixing}

The precise origin of gravity in noncommutative gauge theory is
through a process of \emph{emergent gravity}, as first observed
in~\cite{Rivelles02,Yang06} and clarified in a systematic manner
in~\cite{Steinacker07}. One considers non-vacuum solutions
\beq
\big[X^\mu\,,\,X^\nu\big]\=\ii\theta^{\mu\nu}(x)
\label{nonvacsols}\eeq
of the equations of motion (\ref{eomIKKT}), with position dependent
noncommutativity parameters. Via the Seiberg--Witten map, this
describes a dynamical quantum spacetime. In this setting, gravity is
related to the quantum fluctuations {$X^\mu$} of spacetime at the
Planck scale. Conversely, noncommutative field theory arises as field
dependent fluctuations of spacetime geometry determined by the Poisson
bivector $\theta^{\mu\nu}(x)$ in (\ref{nonvacsols}). This observation
can be used to clarify many of the novel features of noncommutative
gauge theory that we described in \S\ref{gaugeint}. For example, the
{$U(1)$} ``photon'' is really a graviton, which defines a non-trivial
geometric background coupled to {$SU(N)$} gauge fields. This provides
a natural physical explanation for the {$U(1)\leftrightarrow SU(N)$}
entanglement. The resulting gravitational theory is similar to general
relativity for weak curvature. It suggests a new approach to the
quantization and unification of gravity with gauge theory. In
particular, the flat space solution (\ref{Xvacuum}) is stable at
one-loop order.

The formalism also provides a beautiful physical explanation of the
UV/IR mixing problem in noncommutative field theory. For this,
consider the infrared dynamics for momenta $k$ in the regime 
\beq
k<\Lambda<\Lambda_{\rm NC}\=\frac1{\sqrt{\theta}} \ .
\label{1loopregime}\eeq
In~\cite{Grosse08}, the one-loop effective action for noncommutative
gauge fields coupled to dynamical adjoint scalar fields is computed,
involving a careful treatment of the UV/IR mixing terms. By using
well-known heat kernel expansions for scalar field kinetic laplacian
operator, one arrives at an induced Einstein--Hilbert action of the
form
\beq
S\=\int\,\sqrt{g}\,\big(c_1\,\Lambda^4+c_2\,\Lambda^2\,R(g)+
O(\log\Lambda)\big) \ .
\label{EHactionind}\eeq
Thus UV/IR mixing gives a non-renormalizable gravitational sector,
with the ultraviolet cutoff {$\Lambda$} of the gauge theory related to
the newtonian gravitational constant {$G$} according to
(\ref{EHactionind}). This provides a concrete physical explanation for
the non-renormalizability of noncommutative field theories caused by
UV/IR mixing.

The dipole quanta in the ultraviolet regime are in
this case created by the \emph{open Wilson line operators}
{$\Tr\big(\e^{\ii k\cdot
    X}\big)$}~\cite{VanRaamsdonk01,Armoni01,Rey02}, and the physical
interpretation given in \S\ref{UVIRmixing} agrees with the way in
which these operators are argued to couple to gravitational degrees of
freedom in string theory~\cite{Das00}--\cite{Dhar01}. The process
described here is similar to the mechanism of UV/IR mixing that occurs
on a D-brane in a background {$B$}-field, which is related to
tachyonic instabilities arising from the exchange of closed string
modes in the bulk~\cite{Armoni03,Sarkar05}. It is complementary to the
derivation of the newtonian force law~\cite{Ishibashi00} and
supergravity graviton propagators~\cite{Kitazawa07} from
four-dimensional maximally supersymmetric noncommutative Yang--Mills
theory.

\bigskip

\section{Signatures of spacetime noncommutativity\label{NC-signs}}

\noindent
In this final section we will briefly indicate how spacetime
noncommutativity may be observed or measured explicitly in
experiment. There are many proposals for how this may be achieved,
none of which are entirely conclusive. Of course, much of the issue
concerns the magnitude of the effects of noncommutativity, and the
hope is that there exist physical processes for which the effective
scale of noncommutativity lies within the present day experimental
energy range (see \S\ref{Intro}). Here we shall not attempt an
exhaustive survey, but rather just highlight a selection of ideas
which have been put forward, in order to give a flavour for some of
the main issues involved. In particular, we will connect UV/IR mixing
with some of the results of \S\ref{Spacetimequant}.

\subsection{Violations of Lorentz invariance and causality}

Noncommutative field theories provide examples of both Lorentz
invariance and causality violating models, thus destroying the two
pillars of relativistic quantum field theory. In four spacetime
dimensions, the rank two antisymmetric tensor {$\theta^{\mu\nu}$}
provides a directionality
\beq
\mbf\theta_i\=\epsilon_{ijk}\,\theta^{jk}
\label{thetadirection}\eeq
in space. It follows that noncommutative field theory is not invariant
under rotations or  boosts of localized field configurations within a
fixed observer inertial frame. In string theory, this symmetry
breaking is due to the expectation value of a background supergravity
field. A notion of Lorentz invariance may be recovered through a
twist-deformation of the action of the Lorentz group generators on the
noncommutative fields~\cite{Chaichian04,Chaichian04a}.

There is also no sharply localized light-cone (due to spacetime
uncertainty relations), so causality is violated and signals can
travel faster than the speed of light in quantized spacetime. A
notion of localization of noncommutative quantum fields on
``light-wedges'' has been recently developed
in~\cite{Grosse08a}. Thus the spin-statistics theorem and the CPT
theorem do not (necessarily) hold in noncommutative field theory. We
can now try to find physically measurable processes which would
otherwise be forbidden by the usual rules of relativistic quantum
field theory. If observed in experiment, they would vindicate the need
for noncommutative field theory at high (but measurable) energies.

Let us now give a taste of some of the processes that one may attempt
to analyse. In non-abelian gauge theories, such as noncommutative
deformations of quantum chromodynamics or the standard model, the
S-matrix for some processes violates Lorentz invariance. For example,
the Feynman diagram
$$
\mbox{\includegraphics[width=2.5in]{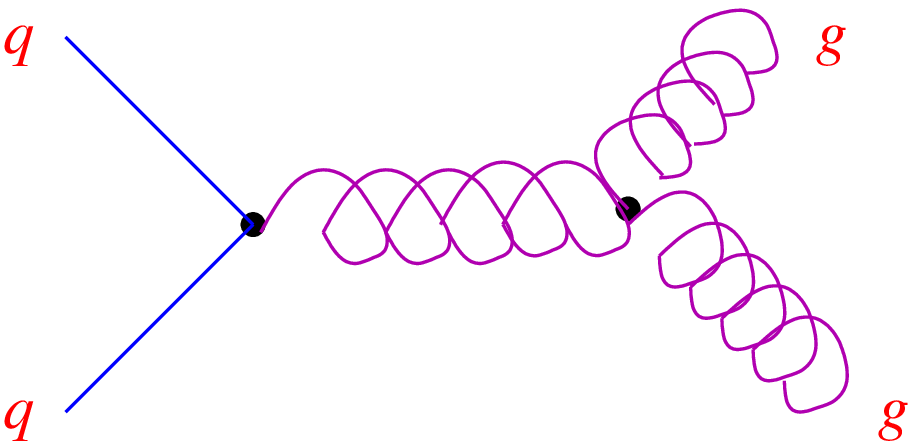}}
$$
is not Lorentz invariant, because the propagator has a
frame dependence on the directionality
$\mbf\theta^0{}_i=\theta^{0i}$~\cite{Balachandran05}. This is
reminescent of Yang's theorem in the commutative case, which forbids
certain diagrams due to Lorentz symmetry violation (despite the
Lorentz invariance of the local 
tree-level lagrangian). However, such processes are allowed in
Lorentz-violating theories such as noncommutative gauge theory. If
even a small amplitude for such a scattering process could be
observed, it would signal evidence for noncommutativity and put bounds
on the magnitude of the parameter $\theta$. For example, a comparison
of the leading terms in a noncommutative deformation of the standard
model with known Lorentz-violating (but CPT symmetric) extensions of
the standard model estimate~\cite{Carroll01}
\beq
\theta\leq\big(10^{-25}~{\rm cm}\big)^{2} \ .
\label{Lorentzvioltheta}\eeq
Other Lorentz-violating processes, which have been studied in the
context of noncommutative extensions of the standard model, include
$Z^0\to\gamma\gamma$~\cite{Buric07,Balachandran07} and decays of
quarkonia~\cite{Tamarit09}.

Violations of causality can be most readily observed by looking at
signatures for spin-statistics violation extracted from atomic
transitions in quantum electrodynamics. One can compare with
experimental results that put limits on the violation of the Pauli
exclusion principle in nucleon systems, based on non-observational
transitions from states which are allowed to occupied levels which are
forbidden by the Pauli principle. For example, comparison with
neutrino data from Gran Sasso or Super-Kamiokande gives a strong
bound~\cite{Balachandran05}
\beq
\theta\leq\big(10^{-24}~{\rm cm}\big)^2 \ .
\label{Paulivioltheta}\eeq
Some simpler analyses of atomic transitions in noncommutative quantum
mechanics can also be carried out, for instance of the Lamb shift in
hydrogen~\cite{Chaichian00}, and the leading corrections due to
noncommutativity compared with experiment.

\subsection{Dispersion relations and UV/IR mixing}

The presence of UV/IR mixing terms in the one-loop effective actions
of noncommutative field theories leads to modified (photon or scalar)
dispersion relations analogous to those discussed in \S\ref{kappaMink}
and \S\ref{3D}. They take the general form
\beq
E^2\=\mbf p^2+m^2+\Delta M^2\Big(\frac1{p\,\theta}\Big) \ .
\label{UVIRdispgen}\eeq
One can compare these dispersion relations with experiments in the
energy range
\beq
\Lambda_0<E<\Lambda\=\frac1{\theta\,\Lambda_0} \ ,
\label{energyrange}\eeq
where {$\Lambda_0$} is the phenomenological infrared scale. For
example, a comparison with observational data from blazars
estimates the noncommutativity parameter
as~\cite{AmelinoCamelia03,Helling07}
\beq
\theta\geq\big(10^{13}\,\ell_{\rm P}\big)^2 \= \big(10^{-20}~{\rm cm}
\big)^2 \ .
\label{blazarest}\eeq
Analogous estimates from cosmic microwave background radiation data
can be found in~\cite{Akofor08}.

\bigskip

\section*{Acknowledgments}

\noindent
The author would like to thank the organisors of the {\sl XXIX
  Encontro Nacional de F\'{\i}sica de Part\'{\i}culas e Campos}, and
in particular V.~Rivelles, for a stimulating meeting and an enjoyable
time in Brasil. This work was supported in part by the
EU-RTN Network Grant MRTN-CT-2004-005104 .

\end{document}